# Detection and Classification of Twitter Users' Opinions on Drought Crises in Iran Using Machine Learning Techniques

**Somayeh Labafi**, Assistant Professor, Iranian Research Institute for Information Science and Technology (IranDoc), Tehran, Iran, labafi@irandoc.ac.ir

**Leila Rabiei**, Iranian Telecommunication Research Center (ITRC), Tehran, Iran, l.rabiei@itrc.ac.ir

**Zeinab Rajabi**, Assistant Professor, Faculty of Computer Department, Hazrat-e Masoumeh University, Qom, Iran, z.rajabi@hmu.ac.ir

**Abstract:**

The main objective of this research is to identify and classify the opinions of Persian-speaking Twitter users related to drought crises in Iran and subsequently develop a model for detecting these opinions on the platform. To achieve this, a model has been developed using machine learning and text mining methods to detect the opinions of Persian-speaking Twitter users regarding the drought issues in Iran. The statistical population for the research included 42,028 drought-related tweets posted over a one-year period. These tweets were extracted from Twitter using keywords related to the drought crises in Iran. Subsequently, a sample of 2,300 tweets was qualitatively analyzed, labeled, categorized, and examined. Next, a four-category classification of users` opinions regarding drought crises and Iranians' resilience to these crises was identified. Based on these four categories, a machine learning model based on logistic regression was trained to predict and detect various opinions in Twitter posts. The developed model exhibits an accuracy of 66.09% and an F-score of 60%, indicating that this model has good performance for detecting Iranian Twitter users' opinions regarding drought crises. The ability to detect opinions regarding drought crises on platforms like Twitter using machine learning methods can intelligently represent the resilience level of the Iranian society in the face of these crises, and inform policymakers in this area about changes in public opinion.

## 1 .Introduction and statement of the problem

The emergence of social media platforms has transformed the structure of human societies. The revolution of social media platforms signifies a transition from the social construction of temporal and spatial reality to the social construction of atemporal and non-spatial reality (Gorska, 2020). (Gurska, 2020). Social media platforms provide the possibility of divergence in societies by networking. Social media users are more likely to be part of a network of social symbols that are in line with their personal beliefs. In social media platforms, networking based on social symbols makes the members of the network more biased. The risk of convergence in the polarization and the fear of social fragmentation has arisen (Sta, 2020). The history of social media platforms in creating social fragmentation in relation to social issues has been studied (Dillon, Neo, & Freelich, 2020). However, some social issues are more important and require further attention. In each period, specific social issues are challenged by the public, and governments prioritize these issues. By examining the current situation in Iran, it can be seen that a social challenge in the context of social resilience against drought has emerged. While there were limited activities around the issue of drought in Iranian society before the rise of social media platforms, the presence of these in response to this issue. Analyzing the theme of resilience in the face of drought provides an analysis of the resilience of Iranians in the face of drought and the behavior of social media platform users in relation to this issue. While no effort has been made so far to understand the issue of resilience in the face of drought on the basis of social media platforms and to use it to predict the behavior of individuals in the real world. Utilizing the Theory of Regulatory focus can shed light on this theme and enable predictions of future behavior. Based on this premise, the aim of this research is to first analyze the content of the Twitter platform in relation to the issue of the drought crisis in Iran using the theory of regulatory focus and classify opinions based on this theory. Then, the development of an intelligent model for the detection and categorization of opinion types in relation to this issue in this platform.platforms has intensified these activities. The risk of social division surrounding the issue of drought can be used as a basis for analyzing the content of social media platforms (Bakar & Hamid, 2020). By using social analysis on platforms, a deep analysis of the reasons and how the behavior of social media platform users is formed in relation to issues such as drought and social resilience in the face of it can be provided (Bronze & Warak, 2020).

Adaptation and Use of Social Science Theories for Analyzing the Content of Social Networking Platforms can clarify the reasons and ways in which users of social networking platforms behave regarding the recent drought issue in Iran. One theory that can be implemented in this context is the theory of regulatory focus. (Pravaz, 2015). The theory of Regulatory focus can be employed to examine the purpose of users from raising the issue of drought and to use it as a basis for predicting their future behavior in the real world.

## 2. Theoretical Concepts of Research

### 2.1. Resilience

Resilience is a broad concept, and the reason for this is the broad implications of this concept. At the individual level, resilience is used to refer to the physical and psychological performance of individuals in dealing with adversity. At the social level, resilience is used to refer to the reflection of the social construction of reality in dealing with adversity. Due to the broad instances, providing a single definition for the concept of resilience is challenging. Resilience can be defined as the ability to improve and recover performance in dealing with adversity (Taylor & Carr, 2021).

The concept of resilience is defined against the concept of inhibitor. An inhibitor is any factor that creates a temporary or persistent disruption to individual and social performance. So far, many classifications have been presented for resilience. In this scheme, three types of resilience, namely physical resilience, mental resilience, and social resilience, are examined (Andia & Chorev, 2021). This classification is comprehensive on the one hand. Comprehensive in the sense that it covers all instances of resilience. On the other hand, this classification is inclusive. Inclusive in the sense that it has the ability to distinguish between different instances of resilience.

### 2.1.1. Physical Resilience

Physical resilience refers to the ability to adapt and maintain the pace of individuals' physical performance in the face of inhibiting factors (Hu et al., 2021). Physical resilience means an individual's endurance in providing the best physical performance when facing inhibiting factors (Liam, 2020). Inhibiting factors in physical resilience refer to any factor preventing the efficient physical abilities of an individual. Inhibiting factors in physical resilience include internal inhibiting factors and external inhibiting factors. Internal inhibiting factors include a range of individual internal factors from any temporary feelings of weakness and deficiency such as oxygen deficiency to any persistent sense of disability such as paralysis of body parts (Colón-Emeric et al., 2020).

External inhibiting factors include a range of external individual factors from any temporary environmental discomfort with temporary consequences, such as environmental temperature, to any environmental discomfort with stable consequences, such as accidents. Usually, internal and external inhibiting factors are related (Riehm et al., 2021). External inhibiting factors affect individual performance when they lead to the emergence of internal inhibiting factors. On this basis, in most cases, external inhibiting factors are the cause of internal inhibiting factors (Sotos-Prieto et al., 2021). For example, high environmental altitude leads to accelerated feelings of oxygen deficiency in individuals, or severe accidents lead to the occurrence of paralysis in body parts. While internal inhibitory factors directly affect an individual's performance, external inhibitory factors require internal inhibitory factors to exert their influence. The role of individual assessment in influencing external inhibitory factors is the reason for presenting the concept of mental resilience.

### 2.1.2.Psychological Resilience

Psychological resilience, also known as psychological well-being, refers to the ability to mentally adapt in order to maintain the pace of mental and physical function of individuals in the face of inhibiting factors (Hu et al., 2021). Psychological resilience means an individual's focus on providing optimal mental and physical performance when facing inhibiting factors (Kilgore et al., 2020). In psychological resilience, inhibitory factors refer to any factor that prevents an individual's concentration in delivering the efficiency of their mental and physical capabilities. These inhibitory factors can be classified into real inhibitory factors and imaginary inhibitory factors. Real inhibiting factors include a range of tangible real inhibiting factors such as fatigue, to intangible real inhibiting factors such as economic pressure (Bozdağ & Ergün, 2020).

On the other hand, imaginary inhibiting factors include a range of individual-specific imaginary inhibiting factors, from not being loved by parents, to culture-specific imaginary inhibiting factors such as favoring death over defeat. Generally, real inhibitory factors and imaginary inhibitory factors are not interrelated. Real inhibiting factors have a noticeable impact on the performance of individuals, and this impact is understandable by others. This is in contrast to imaginary inhibitors that affect the performance of individuals, but this impact is not perceivable by others, either within or outside a cultural context. Individual-specific imaginary inhibiting factors originate from an individual's personal life experience and are not understandable by others, both within and outside a cultural context (Ran et al., 2020). Culture-specific imaginary inhibiting factors stem from an individual's social life experience and are understandable by others within the cultural context, whereas culture-specific imaginary inhibiting factors are not understandable by others outside the cultural context. Imaginary inhibiting factors are stronger than real inhibiting factors. While the selection of individuals in gathering environmental information has a low impact on the evaluation of real inhibitors, the selection of individuals in gathering environmental information plays a major role in the evaluation of imaginary inhibitors. (Hu et al., 2020). Based on this, imaginary inhibitors weaken an individual's ability to accurately collect and evaluate the situation and reduce mental recovery.

Culture-specific imaginary inhibiting factors are more stable than individual-specific imaginary inhibiting factors. The cultural context acts as a barrier against environmental changes. The change in an individual's environment is directly exposed to the individual's view and the individual is free to perceive or not perceive the changes. This is while change in the cultural environment faces cultural resistance (Ran et al., 2020). People in the cultural context easily understand culturally specific imaginary inhibitors. Culture-specific imaginary inhibiting factors are also contagiously transferrable to others (Blanc et al., 2020; Bozdağ & Ergün, 2020 On this basis, individuals within a cultural context easily intensify the influence of cultural imaginary inhibitory factors on others. The level and depth of the impact of cultural imaginary inhibitory factors is the reason for conceptualizing social resilience.

### 2.1.3.Social Resilience

Social resilience or community resilience refers to the ability for mutual adaptation in order to maintain the pace of social and cultural performance of the individuals of society in the face of potentially inhibiting factors (Hu et al., 2021). Social resilience implies the focus of social institutions on providing the best supportive performance when confronting potential inhibitory factors (Berawi, 2020). Social resilience is the most important factor in maintaining social cohesion in conditions of uncertainty. Potentially inhibiting factors in social resilience refer to any potential factor that negatively affects social cohesion by endangering individual and collective interests in society. Inhibiting factors in social resilience refer to any potential factor that prevents focus in providing efficient social and cultural capabilities of the institutions of a country (Jia, Mikami & Normand, 2021).

Not all inhibitors facing human societies are in the face of social resilience. Human societies have a potentially higher ability to cope with inhibitors with the synergy of humans. If the inhibitor is accompanied by certainty (certainty of the direction and intensity of the inhibitor's action), human societies engage in coping with the inhibitor. However, the inhibitor is not always accompanied by certainty. Sometimes the direction of action of the inhibitor is accompanied by uncertainty. For example, the amount of rainfall (the probability of drought or wetness) in the coming years is accompanied by uncertainty. Sometimes the intensity of the inhibitor is accompanied by uncertainty. For example, the amount of drought in the coming years is accompanied by uncertainty. If the inhibitor is accompanied by uncertainty (uncertainty of the direction and intensity of the inhibitor's action), human societies need social resilience in the face of the inhibitor to continue to exist (Garcia & Reim, 2019).

The most important function of social resilience is to reduce the likelihood of social collapse in the face of potential inhibitors. In one classification, the potential inhibitors of human societies were divided into four categories of cultural, social, political, and economic factors (Blanc et al., 2020). In the literature of social resilience, cultural factors include the relations of the human society with other human societies, social factors include inter-institutional relations within the human society, political factors include the relations of institutions with individuals within the human society, and economic factors include interpersonal relations within the human society (Bravai, 2020). However, in recent years, the most important potential inhibitor of human societies has been the factor of climate change. Although climate change has always existed, in recent years the range of direction and intensity of climate change has become more scattered. Based on this, today in the literature of resilience, the potential inhibitors of human societies are introduced in two categories of factors related to climate change and factors unrelated to climate change (Jia, Mckamie, & Normand, 2021). This classification is indicative of the importance of climate change in the literature of social resilience.

**2.1.3.1.Social Resilience Against Drought**

Throughout history, climate change has been the most significant threat to human societies. Climate change threatens the vital resources necessary for individual and social life. At the individual level, climate change leads to a relative decrease in an individual's access to natural resources within a human society. At the social level, climate change creates social tension for greater access of humans to natural resources. Water is the most crucial natural resource for humans. Studies show that more than eighty percent of the global population resides in plain areas (Yudaputra, 2020). These plain regions have direct or indirect access to open seas, inland lakes, or running waters. This indicates the fundamental role of water as a natural resource in shaping and sustaining human communities. Historical studies indicate a relative stability in the ratio of human population living in plains versus plateaus throughout history (Sweya, Wilkinson & Kassenga, 2021).

Climate change does not have an impact on all human societies uniformly. At least two factors moderate the destructive impact of climate change on human societies. The first factor is the climate of a human society. Communities in different climates have diverse experiences. Communities with less access to a specific natural resource become more vulnerable to changes in that resource. Iran's climate makes it more vulnerable concerning water compared to other countries. While over eighty percent of the world's population resides in plain regions, less than twenty percent live in plateau areas (Yudaputra, 2020). In contrast, in Iran, over eighty percent of the population lives in plateau regions, and less than twenty percent in plain areas. Consequently, any change in water resource access, whether due to drought or excess water, will have a severe impact on Iran (Shojaei, Bijani & Abbasi, 2020).

The second factor influencing the impact of climate change on human societies is economic circumstances. Economic conditions vary from underdeveloped to highly developed societies. Underdeveloped societies are more susceptible to climate change, while developed societies are less affected. Developed societies, with better access to the technology required to efficiently manage natural resources, moderate the impact of climate change on their societies. Also, the economies of developed societies rely more on the service sector (Tennberg, Vuojala-Magga & Vola, 2020). In contrast, underdeveloped societies lack the necessary technology to mitigate the impact of climate change on their societies, and their economy relies more on agriculture. Developing societies have limited access to the technology needed to moderate the impact of climate change on their societies, and their economy relies more on industry. Iran, as a developing country, is more vulnerable to climate change compared to developed countries. The focus on industrial development, especially in water-intensive industries, has also led to the impact of changes in access to the natural resource of water in Iran (Fatahi, Vahedi & Arayesh, 2021).

Since the beginning of the 21st century, climate changes in Iran have expanded and intensified drought. While assessments predict even further expansion and intensification of drought in the future. This has resulted in Iran being ranked among the most vulnerable countries to climate change. Iran's vulnerability to climate change phenomena means that Iran generally has limited capabilities to adapt to new climatic conditions, and specifically has limited capabilities to adapt to new drought conditions. Drought is one of the most significant inhibiting factors to social resilience (Ghitrani, Shams & Rahmani, 2018; Hamzekhani, Saghafian & Araghinejad, 2016; Safavi, Shamsai & Saghafian, 2018). Indeed, the threat of drought poses a dual risk in Iran. On one hand, it seriously jeopardizes the well-being and performance of individual Iranians, and on the other hand, it significantly impacts the effectiveness of Iran's institutions and governing bodies.

Based on the latest assessments, on the verge of the 15th century in the Persian calendar, the per capita renewable water availability for Iranian citizens is approximately 1,400 cubic meters. This number is expected to decline to about 13,000 cubic meters on the verge of the 14th century in the Persian calendar. Over a hundred years, per capita renewable water availability for Iranians has decreased by more than 90 percent. Currently, Iran faces a water crisis, with an insecure margin of 36 percent for per capita renewable water availability. The country's management is working towards maintaining access to approximately 900 cubic meters per capita (Hamzekhani, Saghafian & Araghinejad, 2016; Safavi, Shamsai & Saghafian, 2018; Wine, 2020).

## 3. Research Methodology

### 3.1. First phase

The statistical population of the first phase of this research encompasses the entire Persian-language posts related to drought crises in Iran published on the Twitter platform within a one-year timeframe. The main keywords for data collection include water, drought, flood, subsidence, sinkholes, desertification, wastewater, Zayandeh Rud, Karun, Karkheh, Bakhtegan, Zarivar, and lakes. A total of 42.028 tweets were collected and sorted based on four criteria: the number of likes, the number of comments, the number of retweets, and the influence score. After removing duplicate tweets, 25,961 tweets remained.

The sampling method in this study is non-probabilistic and purposive. A purposive approach was used to select indicative tweets, and the criteria for identifying indicative tweets included the influence score, reposting, responses, and likes. The influence score was defined based on a formula of one unit for the number of likes, two units for the number of comments, plus three units for the number of retweets. Based on this formula, the minimum influence score for tweet selection was considered to be 600. The reason for this is to consider a minimum of one hundred for the number of likes, the number of comments, and the number of retweets. Based on this formula, the highly influential tweets collected in this study amounted to 2,935 tweets. The frequency of the collected highly influential tweets is shown in the table 1.

**Table 1: The frequency of influential tweets .**

| Keyword | Number |
|---|---|
| Water | 1774 |
| Drought | 135 |
| Flood | 365 |
| Subsidence | 33 |
| Sinkhole | 7 |
| Desertification | 0 |
| Wastewater | 190 |
| Zayanderud | 45 |
| Karun | 243 |
| Karkheh | 68 |
| Bakhtegan | 5 |
| Zarivar | 3 |
| Lake | 67 |
| Total | 2935 |

After identifying the influential tweets, some of them were excluded because they were in non-Persian languages such as Arabic, Urdu, Punjabi, Pashto, etc. Additionally, tweets without likes, comments, or retweets were also deleted, or if tweets were semantically unrelated to the drought domain, they were also excluded. Finally, a total of 1,142 tweets were considered as the research sample. These research sample tweets were published by 957 distinct user accounts.

### 3.1.1.Data Classification

Next, tweets were analyzed based on the regulatory focus theory. Firstly, based on this theory, the sample tweets were categorized into two groups: proactive tweets and preventive tweets according to the assumption of resilience or lack of resilience against drought. Proactive tweets were assumed to be resilient to drought, while preventive tweets were assumed to be non-resilient to drought. Accordingly, 493 proactive tweets and 649 preventive tweets were identified. The frequency of proactive and preventive tweets can be observed in table 2.

**Table 2: Frequency of Proactive and Preventive Tweets**

| Category | Number |
|----------|--------|
| Proactive | 493 |
| Preventive | 649 |
| Total | 1142 |

Then, based on the regulatory focus theory, preventive tweets were divided into two subcategories gain tweets and non-gain tweets. At the same time, preventive tweets were divided into two subcategories: non-losses tweets and losses tweets. Gain and non-gain tweets were published with the assumption of resilience against drought. In gain tweets, the goal of the publisher is to defend the government's performance in water management for resilience against drought. In non-gain tweets, the publisher relatively defends the government's performance in water management but expresses concerns about society's lack of resilience to drought. Non-losses and losses tweets were published with the assumption of a lack of society's resilience against drought. In non-losses tweets, the goal of the publisher is to criticize the government's performance in water management but maintain hope for resilience against drought. In losses tweets, the publisher criticizes the government's performance in water management and expresses non-resilience to drought. In the preventive category, 415 gain tweets and 78 non-gain tweets were identified. In the preventive category, 137 non-losses tweets and 512 losses tweets were identified.

The thematic analysis method was employed to analyze the data (tweets) in the first phase. After the initial coding of the tweets, the collected data was input into MAXQDA software. Through repeated reading and analysis of the text, initial thematic codes were formed. At this stage, data that were explicitly present in the text or carried implicit meanings within the text were recognized and classified into thematic codes. In this phase, by reviewing and analyzing the collected data, 4,273 thematic codes were created. From these initial thematic codes, 1,472 thematic codes were identified for the gain subcategory, 641 thematic codes for the non-gain subcategory, 769 codes for the non-losses subcategory, and 1,391 codes for the losses subcategory. The frequency of thematic codes is presented in the table 3.

**Table 3: Frequency of identified thematic codes**

| Category | Subcategory | Number |
|---|---|---|
| Proactive | Gain | 1472 |
| | Non-gain | 641 |
| Preventive | Non-losses | 769 |
| | Losses | 1391 |
| Total | | 4273 |

Next, the identified thematic codes were sorted into selection codes based on their semantic similarities. In this phase, it was essential to have a good command of the common language used in tweets. Given the consistency of the native language of the coders with the language of the tweets, the data coding process was carried out with precision.

To maintain research validity, an external control strategy was employed by having an expert third party oversee the researcher's work. To ensure the reliability of the research, two experienced professors in the subject area and research methodology served as consultants and posed challenging questions regarding various aspects of the research design. This included the selection of each component of the research plan, the semantic framework encompassing the choice of subcategories, categories, dimensions, and the interpretive context including the analysis of identified meanings. Additionally, a test-retest reliability method was used to examine the reliability. Test-retest reliability is the percentage of similarity between initial codes identified by a coder at two different times on the same text. In this study, two days after the initial coding of the tweets, 100 tweets were randomly selected and recoded. During the two stages of coding on these 100 tweets, by calculating incomplete and duplicate codes, a total of 211 codes were identified, of which 104 codes were identified as duplicates. According to the test-retest reliability formula, the calculated reliability percentage exceeded 98%, indicating high reliability in the coding process.

$$\text{Reliability Percentage} = \frac{\mathbf{2} * \text{duplicate codes}}{\text{Total codes}} = \frac{2 * 104}{211} = 0.98$$

### 3.1.2. Findings of the first phase:

After analyzing and categorizing the data in various stages of thematic analysis, a total of 85 sub-themes were identified 19 sub-themes were identified for the gain subcategory, 23 sub-themes for the non-gain subcategory, 22 sub-themes for the non-losses subcategory, and 21 sub-themes for the losses subcategory. In the next stage, these sub-themes were combined into the main themes and dimensions of the research. Table 4 shows part of the themes, Sub-themes, and dimensions in the categorization of non-losses tweets, as an example.

**Table 4: Dimensions, Theme, and Sub-themes of the Non-losses tweets category**

| Category | Dimension | Theme | Sub-themes |
|---|---|---|---|
| Non-Losses | Changes in society | Demographic changes | Population growth is a cause of drought |
| | | | Labor force influx is a cause of the drought |
| | | | Rural-to-urban migration is a cause of drought |
| | | | Urbanization is a cause of drought. |
| | | Economic changes | Increasing public demand is a cause of the drought |
| | | | Consumerism is a cause of drought. |
| | | Governance changes | Political decisions in the field of water management are the cause of drought |
| | | | Water-intensive industries are a cause of drought. |
| | | | Unsustainable agricultural development is a cause of drought |
| | | | Unsustainable job creation is a cause of drought. |
| | | | Unbalanced development is the cause of drought |
| | Economic Crisis | Livelihood crisis | Drought is a cause of unemployment |
| | | | Drought is the cause of economic problems |
| | | | Drought is a cause of urban sprawl. |
| | | | Drought is a cause of social inequality. |
| | | crisis of hope | Drought is the cause of despair for the country's future |
| | | | Drought is a cause of emigration. |
| | Changing the consumption pattern | Today's Savings | Water conservation is a solution to drought. |
| | | | Reducing agricultural consumption is a solution to drought. |
| | | | Sustainable use of natural resources is a solution to drought. |
| | | Future planning | Sustainable agricultural development is a solution to deal with drought |
| | | | sustainable industry development is a solution to deal with drought |

## 3.2. The second phase

In this step, keywords frequently occurred in the primary dataset and the conceptual dimensions extracted from the thematic analysis were added to the keyword set presented in Table 1. Subsequently, new data for annotation were extracted from Twitter using a newly prepared keyword set. Then 1258 new labeled data were added to the primary dataset, bringing the total dataset count to 2,300. In the following section, the opinion detection model was trained on this newly constructed dataset.

### 3.2.1.Classification model

In this section, machine learning algorithms were used to identify four opinions related to drought and water crises (namely gain, non-gain, losses, and non-losses) from tweets, and a new classification model was presented. The designed model has the potential to predict drought opinions in new tweets. The following section describes how to design the model and its steps. Refer to Figure 1 for a visualization of the proposed classification model.

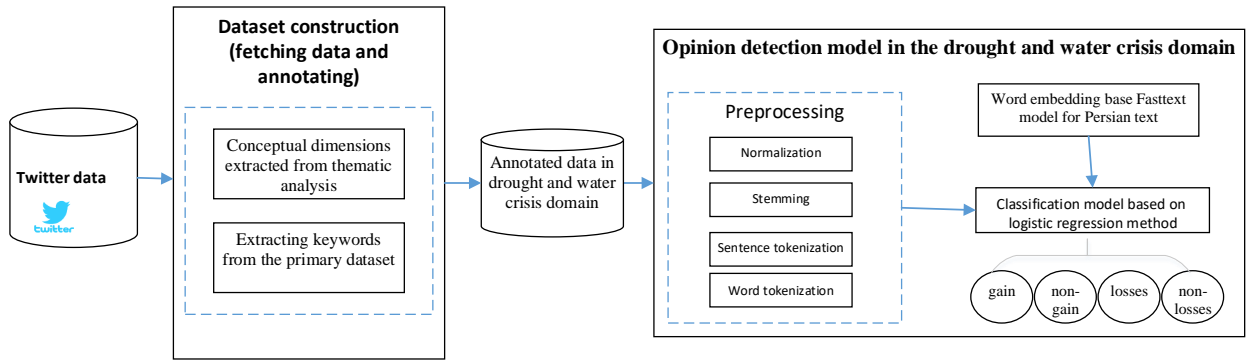

**Figure 1:  Proposed classification model for opinion detection (gain, non-gain, losses, and non-losses) using Twitter posts**

### 3.2.1.1.Word embedding

Many natural language processing models rely on word embedding coding to represent textual data (Zhang, L., S. Wang, and B. Liu, 2018). Typically, word embedding is a technique that transforms words into real-valued vectors, encoding word meanings in a way that proximity in the vector space reflects semantic similarity. Each dimension in the vector represents a hidden feature of a word. These vectors encode linguistic patterns and rules, providing semantic and syntactic information to machine learning models.

that encodes the meaning of the word in such a way that words that are closer in the vector space are expected to be similar in meaning. Textual data should be encoded before constructing a classification model. Word embedding vectors offer better results compared to other encoding methods such as tf-idf. The most common word embedding methods are unsupervised learning methods, which aim to capture linguistic knowledge. Word embedding methods are trained using co-occurrence information and are usually created in general or pre-trained word vectors.

Moreover, natural language processing models can capture the context of a word using word embedding more carefully and encounter less ambiguity (Rajabi, Valavi et al. 2020).

Word2Vec is one of the common word embedding methods, which is actually a computational neural network prediction method learning word embedding from text (Mikolov, Chen et al. 2013). Another word embedding method used is GloVe trained from non-zero entries of a word-word co-occurrence matrix (Pennington, Socher et al. 2014). From a deep learning perspective, semantic vectors or word embeddings allow us to numerically compare the meanings of words; hence, they can be directly used as a representation of word meanings or a starting point for machine learning. Fasttext, a library for learning word embeddings and text classification developed by Facebook's AI Research (FAIR) lab.[. In (Dharma, Gaol et al. 2022), has shown superior results in some cases, compared to the three embedding models for text classification. Persian text has exclusive features and constraints as such it requires a specific language model (Rajabi and Valavi 2021). Accordingly, in this study, we use pre-trained word embedding vectors based on Fasttext models. The opinion detection model for tweets uses Fasttext word embedding for Persian text, which has proven to provide better results than other models in classification.

### 3.2.1.2. Classification

The problem of drought opinion detection is translated into classifying Twitter posts into four categories. Various classification methods, including Naive Bayesian (NB), Support Vector Machine (SVM), Decision Tree (DT), Random Forest (RF), and Logistic Regression (LR), have yielded good results in different applications (Alzanin, Azmi et al. 2022). Moreover, deep neural networks excel in classification but require large datasets for training (Mazoochi, Rabiei et al. 2023, Rabiei, Rahmani et al. 2023). This paper is the first study investigating drought and water crisis in Persian and creates the first relevant dataset.

In the context of subject classification of Twitter posts, the logistic regression classifier has demonstrated successful performance. Among several studies, (Shah, Patel et al. 2020) used the logistic regression classifier for subject classification and reported its successful performance. In this study, the logistic regression classification method was used to classify drought tweets. Since deep neural network classification requires a big dataset, the logistic regression method compared to deep neural network classification is more preferred in this study.

### 3.2.1.3. Evaluation

To evaluate the proposed model, the new dataset of 2,300 tweets was divided into training (80%) and testing (20%) subsets. Data preprocessing tasks, including punctuation correction, word correction, spelling correction, abbreviation insertion, and sentence simplification, were performed to improve sentence detection. The classification based on testing data was analyzed, and the results were reported in Table 5. Multiple experiments and different executions were conducted to classify the data with different parameters of the logistic regression methods, and the best parameters were selected, including tol=0.1, solver=newton-cg, penalty=l2, max_iter=100, C=0.9474722736821756.

**Table 5: Results of classification model for opinion detection**

| F1-score (Micro) | F1-score (Macro) | Accuracy (Test) | Accuracy (Train) |
|---|---|---|---|
| %66 | %66 | 66,09% | 67,77% |

Table 6 shows the Precision, Recall, and F1-score measures for each category: gain, non-gain, losses, and non-losses. As shown, it achieved good results in all categories.

**Table 6: Precision, Recall, and F1-score measures for each category (gain, non-gain, losses, and non-losses)**

| Classes | Precision | Recall | F1-score |
|---|---|---|---|
| gain | 0.74 | 0.66 | 0.70 |
| non-gain | 0.69 | 0.67 | 0.68 |
| non-losses | 0.59 | 0.64 | 0.61 |
| losses | 0.61 | 0.64 | 0.63 |

## 4. Discussion and Conclusion

Adapting social science theories to the context of social network platforms allows for the presentation of analyses aimed at understanding the reasons and ways of changing the behaviors of users on social network platforms. Consequently, the current research leveraged the regulatory focus theory to extract opinion categories present in the content generated by users on the Twitter platform in the context of resilience against drought. In this context, the regulatory focus theory provides a basis for comprehending the users' motives for producing and publishing content related to drought resilience on the Twitter platform. This study represents the initial exploration of analyzing public opinions on drought and water crises in Iran using social media data. By utilizing the capabilities of text mining and machine learning in the field of data analysis, a novel model has been introduced for identifying opinion categories within the subject of drought. The presented model is based on the word embedding model and employs logistic regression classification. This model can identify and predict different opinion categories related to drought on Twitter with a high degree of accuracy.

In this research, based on the regulatory focus theory, four distinct opinion categories within the domain of drought crises in Iran have been identified, followed by the design of an intelligent model for detecting these opinions on Twitter. The regulatory focus theory posits that humans possess an internal supervisory system that guides all their behaviors, categorizing them into two categories promotion and prevention. A concentration on achieving positive goals characterizes promotion-focused behaviors. In this sense, any kind of human behavior whose subject is considered positive by humans can be called promotion focus. Accordingly, if users have a positive perspective and experience regarding the government's performance in water management and take actions to encourage resilience against drought, then their content production behavior is shaped under the gain opinion category. All tweets with this purpose have been extracted and analyzed, forming the first opinion category: " the gain category."

According to the regulatory focus theory, if an individual considers a subject positively but does not take action to defend it, this leads to non-gain behaviors. In such cases, the person deprives himself of the benefits of the subject due to reasons such as social pressures, resulting in a lack of alignment between their opinions and their actions. Within this framework, tweets produced by users who expressed positive opinions about the government's performance in water management but did not directly defend or made no effort to expand on those views were categorized as non-gain opinion category. The non-gain opinion category can be further explained through four dimensions: societal changes, inefficient management, identity crisis, and existential crisis.

Within the regulatory focus theory, the second system is prevention focus. In the prevention focus system, individuals have a negative perception of the subject. In this category, the subject is seen as having negative qualities, and individuals may aim to avoid the negative aspects of the subject. by three themes governance changes, demographic changes, and economic changes. The economic crisis dimension is explained by two themes hope crisis and livelihood crisis. Finally, the changing consumption pattern dimension is also explained by two themes of saving today and planning tomorrow. In the prevention focus system, individuals have a negative perception of the subject. In this context, governments have poor performance in water management. This group of users pursues direct confrontation and explicit criticism against it. On this basis, they produce and publish content against the government's performance.

In conclusion, this research introduces several innovations, including:

- Using the regulatory focus theory as a foundational framework for understanding and categorizing user opinions on the Twitter platform.
- Creating a new dataset consisting of 2300 labeled tweets related to drought and water crisis in Iran.
- Formulating the problem of analyzing the opinions of the Iranian society regarding the water crisis in Iran as a machine-learning and classification problem for the first time
- Presenting a new model for classifying and detecting opinions within the water crisis domain, based on word embedding and logistic regression classification, specifically for analyzing Persian-language Twitter users.